# Measurement and modelling of anomalous polarity pulses in a multi-electrode diamond detector

J. Forneris[1], V. Grilj[2], M. Jakšić[2], P. Olivero[1], F. Picollo[1], N. Skukan[2], C. Verona[3], G. Verona-Rinati[3], E. Vittone[1]

[1] Physics Dept., NIS Centre and CNISM, Torino University; INFN sez. Torino; via P. Giuria 1, 10125 Torino, Italy
[2] Ruđer Bošković Institute, Bijenička cesta 54, P.O. Box 180, 10002 Zagreb, Croatia
[3] Dipartimento di Ingegneria Industriale, Università di Roma "Tor Vergata", Via del Politecnico 1, 00133 Roma, Italy

**Abstract –** In multi-electrode detectors, the motion of excess carriers generated by ionizing radiation induces charge pulses at the electrodes, whose intensities and polarities depend on the geometrical, electrostatic and carriers' transport properties of the device. The resulting charge sharing effects may lead to bipolar currents, pulse height defects and anomalous polarity signals affecting the response of the device to ionizing radiation. This latter effect has recently attracted attention in commonly used detector materials, but different interpretations have been suggested, depending on the material, the geometry of the device and the nature of the ionizing radiation. In this letter, we report on the investigation in the formation of anomalous polarity pulses in a multi-electrode diamond detector with buried graphitic electrodes. In particular, we propose a purely electrostatic model based on the Shockley-Ramo-Gunn theory, providing a satisfactory description of anomalous pulses observed in charge collection efficiency maps measured by means of Ion Beam Induced Charge (IBIC) microscopy, and suitable for a general application in multi-electrode devices and detectors.

**Introduction** – Solid-state multi-electrode detectors are widespread devices for radiation measurement. Segmented devices such as strip, drift, pixelated detectors display advantageous features such as high spatial resolution, low capacitance, low noise, thus matching the requests for high reliability, reproducibility and spectral resolution for radiation tracking and measurement in particle physics, photodetection, medical imaging and dosimetry [1]. On the other hand, in a multi-electrode device, the motion of free carriers generated by ionizing radiation induces a charge on the electrodes surrounding the active region. The sharing of the induced charge depends on the device geometry and the electronic properties of the material, and may lead to to possible misinterpretations of measured signals, if compared with the response of detectors with large surface electrodes. Moreover, a solid understanding of charge sharing mechanism in multi-electrode devices provides an essential tool for the development of promising deterministic doping techniques (i.e. position resolved single ion implantation) for quantum computing applications [2,3].
In the past decades, several limitations to the sensitivity of multi-electrode detectors were ascribed to charge sharing in commonly used silicon, cadmium zinc telluride (CZT) and germanium segmented detectors. Pulse height defects and charge losses [1], charge collection efficiency (CCE) losses and spectra distortion [4-6], ambipolar transient current signals [7] and anomalous polarity pulses [8,9] have been reported so far. Specifically, we adopt the "anomalous polarity pulse" term to refer to an induced charge pulse having the opposite polarity to what would be detected in a standard device with two large parallel electrodes [9].
A few studies [1,8,9] proposed the Shockley-Ramo theory [10,11] and its extensions at the basis of the interpretation of such phenomena, and identified the occurring of anomalous polarity pulses as a key fingerprint of charge sharing in strip detector geometry [8,9].
As the maturity of the above-mentioned materials for radiation detection applications ensures a complete charge collection at the electrodes [5,7,12], charge sharing and related effects were naturally ascribed to carriers thermal diffusion in inter-electrode gaps [4,5,13,14] and to an extended charge cloud size [5,12,14,15].
Moreover, charge losses and anomalous polarity pulses were attributed to carriers recombination [14], to a modification of the electric field and the charge collection geometries by the presence of charges trapped at the surface [1,8,9] or by radiation damage effects [1].

*Corresp. author: Jacopo Forneris, Physics Department, University of Torino - via P. Giuria 1, 10125 Torino, Italy. E-mail: jacopo.forneris@unito.it

In this work, we report on the first observation of anomalous polarity pulses in diamond and on their interpretation, based on a pure electrostatic model.

Diamond was chosen as a benchmark material for two main reasons. Firstly, the development of multi-electrode geometries in diamond detectors is highly desirable because of its promising features such as high radiation hardness, high carrier saturation velocities and low thermal noise. With this purpose, an assessment of the impact of charge sharing effects on multi-electrode diamond detectors is demanded. Secondly, the carriers lifetimes significantly shorter than in average traditional materials such as CZT, germanium, silicon enable a full investigation in the effects of carriers recombination in the bulk on the formation of anomalous polarity pulses without the request of *ad hoc* constraints on the charge collection geometry associated with irradiation, inhomogeneous trap distribution or charge accumulation at the oxide. Thus, the study of the bulk recombination processes in diamond enables us to adopt and validate a general model based on a purely electrostatic approach and relying on the Shockley-Ramo-Gunn (SRGT) theorem [16,17]. The generality of the latter will allow a straightforward application to multi-electrode devices in any detector material, and a further corroboration of results in [1,8,9].

The experimental measurements were carried with the Ion Beam Induced Charge (IBIC) technique, which allowed the microscopic mapping of the induced pulse height as a function of the excess charge-carriers generation position across the device [16].

**Theory** – A model for the evaluation of the current instantaneously induced at the *j*-th electrode by a point-like charge $q$ moving in a system with an arbitrary arrangement of $n$ conductors and space charges is given, under general assumptions, by SRGT [16,17]:

$$i_j = \frac{d(Q_{j,0}(\mathbf{x}(t)) - Q_{j,0})}{dt} = -q\,\mathbf{v} \cdot \frac{\partial \mathbf{E}}{\partial V_j} = +q\mathbf{v} \cdot \nabla\left(\frac{\partial \psi}{\partial V_j}\right) = q\mathbf{v} \cdot \nabla \varphi_j = -q\mathbf{v} \cdot \mathcal{E}_j \quad (1)$$

where **v** is the velocity vector of the moving charge, $\psi$ is the electrostatic potential and $\varphi_j = \partial \psi/\partial V_j$ is the Gunn's weighting potential, defined as the derivative of the electric potential with respect to the voltage $V_j$ applied at the *j*-th electrode; similarly, the weighting field $\mathcal{E}_j = \partial \mathbf{E}/\partial V_j$ is given by the derivative of the electric field **E** with respect to $V_j$. In the first equality in Eq. (1), we defined the charge induced at the sensitive electrode as the difference between the total charge $Q_j$ induced at the *j*-th electrode when $q$ is at the position **x** at time $t$, and the charge $Q_{j,0}$ stored at the *j*-th electrode when the system is at the electrostatic equilibrium (i.e. in absence of $q$). In absence of a voltage-dependent internal space charge distribution, Eq. (1) reduces to the usual Shockley-Ramo theorem [16].

The integration of Eq. (1) over a time interval $\Delta t = (t_f - t_i)$ leads to an expression for the charge induced at the *j*-th sensitive electrode by the motion of the point charge $q$ from the initial position $\mathbf{x}_i = \mathbf{x}(t_i)$ to the position at a final time $t_f$, $\mathbf{x}_f = \mathbf{x}(t_f)$ [18]:

$$q_j = \int_{t_i}^{t_f} \frac{d(Q_{j,0}(\mathbf{x}(t)) - Q_{j,0})}{dt} dt = q[\varphi_j(x_f) - \varphi_j(x_i)] \quad (2)$$

The formalism adopted in eq. (2) is suitable for a direct application in multi-electrode devices, clarifying the nature of the induced charge polarity at each electrode.

In fact, by the charge conservation principle, the total excess charge in the device, at any time *t* must be equal to the injected charge $q$. As a consequence, the sum of the charges $q_j$ induced by $q$ over all the *N* electrodes of the device is identically null:

$$\sum_{j=1}^{N} q_j = \sum_{j=1}^{N}(Q_{j,0}(\mathbf{x}(t)) - Q_{j,0}) = 0, \quad \forall t \quad (3)$$

Which states that the sum of the total induced charge over the electrode index is zero, at any instant of the carrier motion within the system. Moreover, from Eq. (2),

$$\sum_{j=1}^{N}\left[\varphi_j(x_f) - \varphi_j(x_i)\right] = 0, \quad \forall x_f, x_i$$

(4)

If the charge is located at the *k*-th electrode (i.e. $\mathbf{x}=\mathbf{x}_k$), the following identity holds: $\varphi_j(x_k) = \delta_{jk}$ ($\delta$ is the Kronecker delta symbol) and, hence, for every point of the device the sum of the weighting potentials is equal to one:

$$\sum_{j=1}^{N} \frac{\partial \psi(\mathbf{x})}{\partial V_j} = 1, \forall \mathbf{x}$$

(5)

Eqs. (4-5) are particularly convenient in order to investigate the charge induction mechanism associated with a point-like carrier in motion, i.e. to study charge sharing effects without the assumption of and extended carriers cloud size [5,12,14,15].

A straightforward consequence of Eqs. (4-5) is the existence of a natural definition for an "intrinsic" charge sharing effect, referred to as the phenomenon in which the same carrier concurrently induces charge at more than one electrode, due to its motion in the system domain. We believe that such a definition allows disambiguating among the different interpretations proposed for the arising of charge sharing phenomena and referring to the spot size of the ionizing radiation, the size and the spreading of the charge pair cloud and thermal diffusion of carriers in the inter-strip region of segmented detectors.

A first simple result of Eqs. (5) is that charge is always induced concurrently at least on two electrodes. Particularly, the condition of the overlapping of weighting potentials (sometimes referred to as "weighting potential crosstalking" [7]) is always true in the case of a two-electrode system. However, it is easy to show that the electric field lines of any two-electrode system are topologically equivalent to a one-dimensional geometry (e.g., a parallel-plate capacitor), in which the weighting field is parallel to the electric field through the whole electrostatic domain. Therefore in this case, since the induced charge is given by Eq. (2), both electrodes bare the same amount of induced charge on their plates, the only difference being the opposite relative sign in charge polarity, which is unequivocally defined by the carriers drift velocity orientation (either parallel or anti-parallel to the electric field lines).

On the other hand, a non-trivial "intrinsic" charge sharing occurs when one charge carrier *q* crosses a region exhibiting a crosstalking in the weighting potentials of three or more electrodes. In this case, the global charge conservation condition in Eq. (5) implies a redistribution of the total charges induced at the electrodes, and therefore it determines their polarity.

We define the induced signal polarity at the *j*-th electrode as the sign of $q_j$, given by the product of the sign of the charge *q* times the difference in weighting potential between its final and initial positions (Eq. 2).

Differently from the electric field topology of two-electrode devices, in a multi-electrode configuration the carriers drift velocity may assume different orientations with respect to the weighting field; as a consequence, the sign of the instantaneous induced current (Eq. (1)), and hence the charge pulse polarity (Eq. (2)), can vary depending on the electrostatic configuration of the region crossed by the charge [17,18,19].

Let us discuss the polarity of induced charge pulses in the one-sided planar strip geometry semiconductor device with four parallel electrodes (I-IV) represented in Fig. 1, in which the two central electrodes are grounded and the two external are biased to -100 V.

It is worth noting that the geometry under investigation is equivalent, from the point of view of the electric field topology, to many detectors with segmented electrodes in silicon [1,3,9] and CZT [7,15]. In fact, despite in such works the electric field is mostly perpendicular to the top active area of the detector, the excess charge carriers drift along the electric field lines, which can be continuously deformed into the configuration shown in Fig. 1. Therefore, given that the proposed planar geometry does not represent a regular detector, it is however fully equivalent to commonly adopted segmented detectors from the point of view of the charge transport geometry and the role of carrier species in the pulse formation mechanism.

Assuming the absence of any space charge, the electrostatics of the device is described by the Laplace's equation. A Finite Element Method (FEM) evaluation of the electric field and potential associated with the device, implemented according to the procedure described in [20] is shown in Fig. 1a. The two sub-systems (i.e., the I-II and III-IV electrodes) act as separate capacitors in a mirror-like configuration with respect to the axis of symmetry of the system and their electric field lines do not intersect, meaning that a positive (or negative)

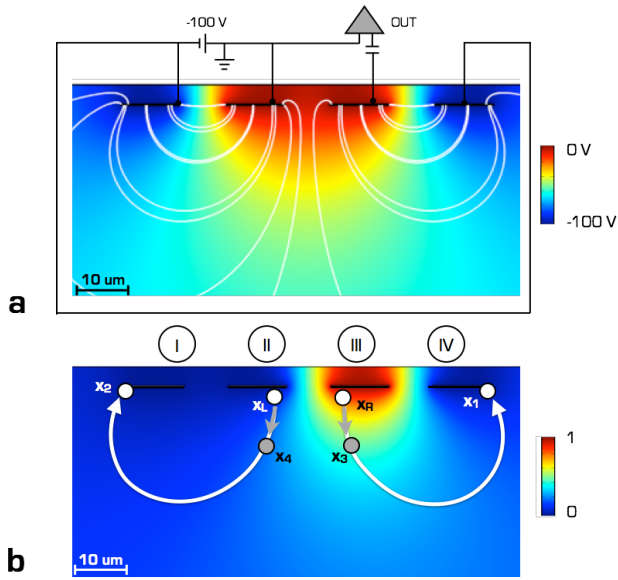

**Figure 1**: FEM two-dimensional electrostatic model of the sample under investigation. The voltage configuration is sketched in (a), together with the electric potential map and the electric field streamlines (in gray). In (b) the map of the weighting potential associated with electrode III is shown. A sketch of the drift trajectories for a positive charge collected at the electrodes (green arrows) or recombining in the bulk (gray arrows) for different starting positions is represented.

charge drifts towards the closest cathode (or anode, respectively).

The weighting potential map relevant to the sensitive electrode III is shown in Fig. 1b (the weighting potential for electrode II can be obtained as a mirror image). It is worth noting that, differently from the case of an isolated capacitor, the weighting potential map is not symmetric and can assume non-zero values in the opposite semi-plane of the device [1,3]. Therefore, according to Eq. (2), a moving charge can induce a signal on electrode III even if it is generated in the left semi-plane and collected by electrodes I or II.

Let us now discuss the polarity of the charge induced at electrode III when a positive charge $q$ is generated within the device. Two different generation points placed in the left (position $\mathbf{x}_L$) and right ($\mathbf{x}_R$) semi-planes are considered (Fig. 1b).

Let us assume first that $q$ drifts according to the electric field lines and it is collected at the nearest cathode. If $q$ is generated at $\mathbf{x}_R$, it will be collected at electrode IV (position $\mathbf{x}_1$). Since the value of the weighting potential is greater at the generation point than at the collection position, Eq. (2) predicts a negative charge induced at electrode III.

On the other hand, if $q$ is generated at $\mathbf{x}_L$, it will be collected at electrode I (position $\mathbf{x}_2$). Again, since the weighting potential at the final position ($\mathbf{x}_2$) is zero, the charge induced at electrode III can be negative (if $\phi_{III}(\mathbf{x}_L)>0$) or null (if $\phi_{III}(\mathbf{x}_L)=0$). As a general conclusion, the polarity of any measured pulse induced at a sensitive anode by a positive charge $q$ collected at a cathode is always negative or null, i.e. $q \leq 0$.

Let us now consider the case, in which $q$ recombines before collection, due to a short lifetime in the device's material. If $q$ is generated at $\mathbf{x}_R$, it will recombine at a generic point $\mathbf{x}_3$. Since the weighting potential is monotonically decreasing from the anode to the cathode along the field lines, the induced charge will still be negative, i.e. $q \leq 0$. However, if $q$ is generated at $\mathbf{x}_L$, recombination may occur at an alternative position $\mathbf{x}_4$ where the weighting potential relevant to electrode III is higher, i.e. $\phi_{III}(\mathbf{x}_4)>\phi_{III}(\mathbf{x}_L)$. In this case, the induced charge evaluated through Eq. (2) has a positive polarity, i.e. $q \geq 0$. We refer to such a polarity as "anomalous", since the positive charge $q$ is acting as an equivalent negatively-charged carrier in an isolated two-electrodes capacitor [16], as a consequence of the peculiar topology of the internal electric field.

We conclude that a carrier can induce charge with both polarities at the same electrode, depending on the position of the generation point, the geometry of the electric field and the transport properties of the material.

**Experimental** – A previously characterized multi-electrode diamond detector with buried graphitic electrodes [21] was employed to validate the proposed model.

The device under test is based on synthetic single-crystal diamond film grown by plasma-enhanced microwave Chemical Vapor Deposition technique at the University of Rome "Tor Vergata" on a commercial 4×4×0.4 mm³ High Pressure High Temperature (HPHT) Ib single-crystal diamond substrate. The thickness of the diamond layer was approximately 40 $\mu$m.

Buried graphitic electrodes were fabricated with a Deep Ion Beam Lithography (DIBL) process [20]. DIBL was performed using a ∅~10 µm focused 1.8 MeV He beam and raster-scanned along linear paths. The ion fluence (~1.5×10$^{17}$ cm$^{-2}$) was set to induce the formation of amorphous channels at ~3 µm below the sample surface (see Fig. 2a) [22], according to SRIM simulations [23]. The electrical continuity of the buried channels with the sample surface was ensured by the deposition of slowly-thinning Cu masks, resulting in the modulation of the depth of the implanted layers [20]. The sample was then annealed in vacuum at 1100 °C for 2 hours, both to partially recover the ion-induced structural damage in the region overlying the buried channels, and to promote the conversion of the damaged region to a graphitic phase [22].

The diamond sample was micro-fabricated with four ~10 µm wide parallel electrodes, plus an additional orthogonal channel. The spacing between the buried electrodes was ~12 µm. Finally, 80 nm Cr/Al circular contacts (∅=150 µm) were patterned by standard photolithographic technique at the emerging endpoints to wire-bond the electrodes with the external circuitry (optical micrograph in Fig. 2b).

IBIC measurements (see [21] for further details) provided a preliminary characterization of the detector's performance. From the IBIC analysis, hole and electron lifetimes were estimated to have constant values of 0.4 ns and 25 ps in the pristine material, respectively. The significant imbalance between hole and electron lifetimes allows the device to be regarded as "single carrier", and hence suitable to validate the model proposed in Sect. 2.

The electrical configuration reported in Fig. 2b, corresponding to the FEM model in Fig. 1, was adopted to perform an IBIC analysis on the charge pulse polarity. Electrodes II and III were grounded, III being the sensitive electrode. A -100 V voltage was applied to electrodes I and IV. In addition, the horizontal electrode was held at a constant potential of -100 V to prevent effects on the measurement due to the presence of a floating electrode. It is worth noting that such a configuration is not significantly different from that proposed in Fig. 1, as the topology of the electric field is still equivalent.

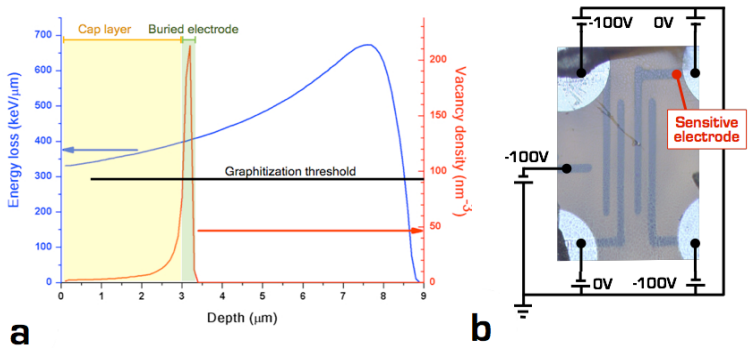

**Figure 2** (a) SRIM simulation of energy loss of 4 MeV He ions (blue line, left vertical scale) and vacancy density profile in diamond generated by 1.8 MeV He ions implanted at a fluence of ~1.5×10$^{17}$ cm$^{-2}$ (red line, right vertical scale) [23]. (b). Optical micrograph of the micro-fabricated structure with the voltage configuration adopted in the IBIC experiment.

The rise of anomalous polarity pulses is expected when electron-hole pairs generation is close to $x_t$, i.e. in the region where the weighting potential is close to zero (within ~5-10 µm in depth, Fig. 1b). IBIC measurements were performed at the ion microbeam line of the Laboratory for Ion Beam Interactions of the Ruđer Bošković Institute (Zagreb) using a ∅~4 µm focused 4 MeV He beam (penetration depth ~9 µm, Fig. 2a). The low-current beam (<10$^3$ ions s$^{-1}$ to prevent pile-up effects) was scanned over the diamond surface. Each incident ion generates a measurable charge pulse, which is amplified and processed by a standard charge-sensitive electronic chain. The data acquisition system stores each event along with the coordinates of the ion incidence. The electronic chain connected to the sensitive electrode featured an Ortec142 charge-sensitive preamplifier and an Ortec570 amplifier (shaping time: 0.5 µs). The pulse heights were calibrated using a reference Si surface barrier detector and a precision pulse generator, resulting in a spectral sensitivity of ~1290 electrons/channel (0.4% CCE). The noise threshold was set to 8% CCE.

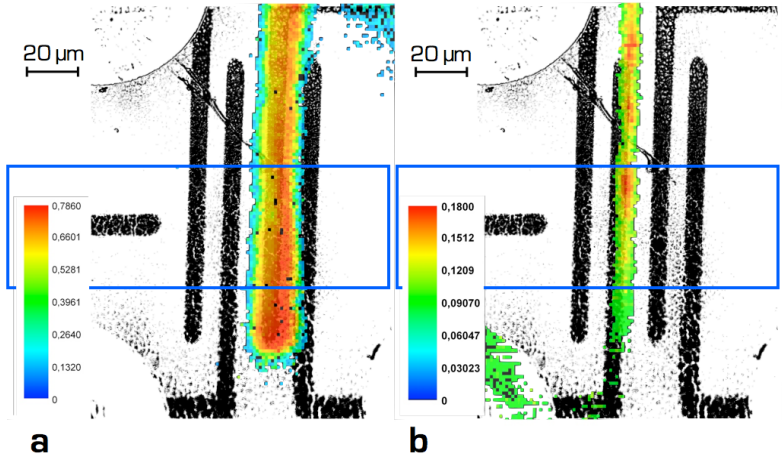

**Figure 3** Positive (a) and negative (b) polarity median CCE maps acquired at electrode III (see Fig. 2b). An optical (b/w) micrograph of the sample is superimposed to identify the relative position of the detected pulses. The blue rectangle highlights the area from which the horizontal profiles in Fig. 4 were extracted.

Since the pulse polarity is inverted by the charge sensitive preamplifier, we firstly selected a positive polarity on the shaping amplifier to collect the "ordinary" signals. The acquired median CCE map is shown in Fig. 3a. The IBIC signals reached a maximum of ~80% CCE, reproducing the shape of the sensitive electrode.
The shaping amplifier was then switched to negative polarity to collect anomalous pulses. By comparing the resulting median CCE map (in Fig. 3b) with the optical micrograph in Fig. 2b, we identified the position of the anomalous polarity pulses as originating from the right edge of electrode II, as expected from schematics in Fig. 1. Pulse heights (up to ~18% CCE) are compatible with the variation in weighting potential along the electric field lines in Fig. 1b. Median CCE profiles along the horizontal axis extracted from the IBIC maps (blue area in
Fig. 3) are shown in Fig. 4a.
A FEM simulation of the experiment was performed. The 2-dimensional electrostatic model in Fig. 1 was used as input; mobility, saturation velocity and lifetime for both carrier species ($\mu_n$=2200 cm$^2$ V$^{-1}$ s$^{-1}$, $\mu_p$=3400 cm$^2$ V$^{-1}$ s$^{-1}$, $v_{n,sat}$=1.35×10$^7$ cm s$^{-1}$, $v_{p,sat}$=1.45×10$^7$ cm s$^{-1}$, $\tau_p$=0.4 ns, $\tau_n$=25 ps) were set consistently with what reported in [21].
The CCE map was then obtained by the FEM implementation of the adjoint equation method [16,24]. The result, shown in Fig. 4b, presents the CCE map as a function of the electron-hole pair generation position, without considering ionization profiles or beam dispersion. Therefore, the negatively-signed CCE around electrode II (cfr. Fig. 1b) indicates that the arising of anomalous polarity pulses is compatible with the model for "intrinsic" point-like charge sharing presented above. In fact, the point-like generation associated with the adjoint equation method stresses that the anomalous polarity effect cannot be explained in terms of the size of the charge cloud generated around the symmetry axis of the system. Additionally, contributions from carrier

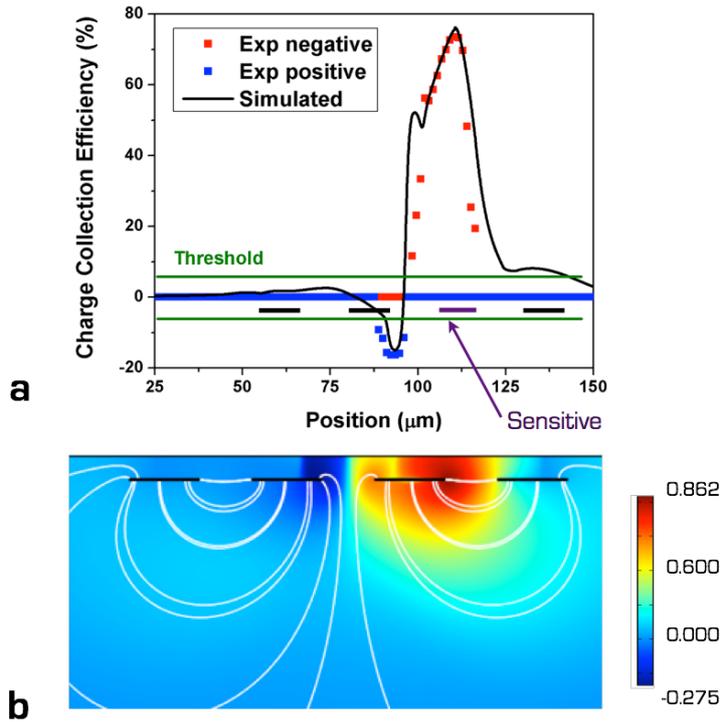

**Figure 4** (a) Ordinary (red dots) and anomalous (blue dots) median CCE profiles along the CCE profile. The two horizontal lines represent the threshold of the acquisition system. (b) FEM-simulated CCE maps. The charge collection efficiency is plotted as a function of the point-like generation position of electrons an holes, according to the solution of the adjoint equation.

diffusion and electron cloud are negligible due to the short carrier lifetimes and to the microbeam spot size, respectively.

It is worth stressing that, despite the distance between electrodes (few tens of $\mu$m) is comparable with the chosen drift lengths, the pathway of carriers drifting along the electric field lines (Fig. 1, Fig. 4b) is significantly longer and results in an incomplete charge collection.

Finally, the convolution of the CCE map with the ionization profile of 4 MeV He ions determines the simulated CCE profile, which is superimposed to experimental data in Fig. 4a, showing a good agreement in terms of peak position and height.

**Conclusions** – In the present work we reported on the observation of anomalous polarity pulses in a diamond detector with multiple buried graphitic electrodes. The proposed model, based on a purely electrostatic approach and relying on the SRGT and on FEM numerical simulations provided a faithful reproduction of the experimental results, in terms of location and height of anomalous polarity pulses.

The model naturally extends the Shockley-Ramo theory to multi-electrode devices, assuming that the anomalous polarity pulses are related with charge-sharing phenomena in multi-electrode devices, as a consequence of the recombination of charge carriers in the bulk, preventing a full collection at the electrodes, in a region of the device where electric and weighting fields are not parallel. Such an interpretation is supported by the study of the CCE as a function of the generation position, which has been performed by neglecting any effect due to the charge cloud size and spread. In addition, differently from the significant contribution to charge sharing phenomena in the case of the collection of carriers at the electrodes [5,7,12], diffusion cannot be regarded as a cause of anomalous polarity pulses due to the short carriers' lifetime in the material under test. The model is in agreement with the interpretation of results found in [1,4,8,9], ascribing the phenomenon to a selective recombination of carriers before collection at the electrodes; however, while in such works anomalous polarity pulses are induced by external factors (e.g. surface trap density or radiation damage effects) requiring additional assumptions on non-homogenous charge trap distributions, in our case they can be regarded an intrinsic effect, due to the charge transport properties of the material under test.

The intrinsic properties of the material under test enabled us to validate the proposed model, whose general formulation can be successfully applied to the interpretation of general multi-electrode devices and detectors. We envisage that this work will provide a useful reference for the development of multi-electrode and 3-dimensional diamond detectors [25], whose exploitation is attracting growing interest due to their enhanced radiation hardness properties in particle physics and medical applications. In these fields, charge sharing can play a relevant role in the response of ionizing radiation detectors, leading to position-dependent effects on the measured pulse polarity.

More generally, we believe that this approach can be successfully applied in the diagnostics and the engineering of bipolar effects in multi-electrode devices, in which an additional effort in the design of electrodes geometry and internal electric field is required to prevent operation limitations.

***

This work was supported by the European Community as an Integrating Activity 'Support of Public and Industrial Research Using Ion Beam Technology (SPIRIT)' under EC contract no. 227012; by INFN experiment DIAMED; by MIUR, PRIN2008 National Project ''Synthetic single crystal diamond dosimeters for application in clinical radiotherapy''; by MIUR, "FIRB - Futuro in Ricerca 2010" project (CUP code: D11J11000450001) and by the University of Torino-Compagnia di San Paolo-projects ORTO11RRT5, 2011- Linea 1A and call "Call1-D15E13000130003". The work of Claudio Verona was supported by "Fondazione Roma", which is gratefully acknowledged.